\documentclass[preprint2]{aastex}

\usepackage{natbib}
\usepackage{amsmath}
\usepackage{booktabs} 
\usepackage{tabularx}

\usepackage{natbib,twoopt}
\usepackage[breaklinks=true]{hyperref} 

\makeatother
\usepackage{graphicx}
\usepackage{txfonts}

\shorttitle{Point and Interval Estimation on Polarization Measurements.}
\shortauthors{Maier et al.}

\begin{document}

   \title{\vspace{-7mm}Point and Interval Estimation on the \\Degree and the Angle of Polarization.\\
    -- A Bayesian approach --}
	
   \author{\vspace{-5mm}{D. Maier\altaffilmark{*}} and C. Tenzer, A. Santangelo}
   \affil{Institute for Astronomy and Astrophysics T\"ubingen (IAAT),
       		Sand 1, 72076 T\"ubingen, Germany\\
       		\vspace{2mm}
       		\textnormal{Source: Publications of the Astronomical Society of the Pacific, Vol. 126, No. 939 (May 2014),
pp. 459-468\\
       		Stable URL: \url{http://www.jstor.org/stable/10.1086/676820}\\
       		\copyright\,2014 by the Astronomical Society of the Pacific.
       		\vspace{-5mm}
         	}}
   \altaffiltext{*}{e-mail: maier@astro.uni-tuebingen.de}

\begin{abstract}
Linear polarization measurements provide access to two quantities, the degree (DOP) and the angle of polarization (AOP). 
The aim of this work is to give a complete and concise overview of how to analyze polarimetric measurements. We review interval estimations for the DOP with a frequentist and a Bayesian approach. Point estimations for the DOP and interval estimations for the AOP are further investigated with a Bayesian approach to match observational needs.
Point and interval estimations are calculated numerically for frequentist and Bayesian statistics. Monte Carlo simulations are performed to clarify the meaning of the calculations.
  
Under observational conditions, the true DOP and AOP are unknown, so that classical statistical considerations -- based on true values -- are not directly usable. In contrast, Bayesian statistics handles unknown true values very well and produces point and interval estimations for DOP and AOP, directly. Using a Bayesian approach, we show how to choose DOP point estimations based on the measured signal-to-noise ratio. Interval estimations for the DOP show great differences in the limit of low signal-to-noise ratios between the classical and Bayesian approach. AOP interval estimations that are based on observational data are presented for the first time. All results are directly usable via plots and parametric fits.
\end{abstract}
   

%
\section{Introduction}
Polarization of electromagnetic waves can be described with the concept of elliptical polarization, which implies linear and circular polarization as special cases. In this paper, the term 'polarization' is used to refer to linear polarization. 
Polarization measurements provide access to two quantities, the degree of polarization, which describes the relative amount of polarized photons to all observed photons, 
and the angle of polarization, which gives the orientation of the electric field of the electromagnetic wave.
Point and interval estimations can be performed on both observables with the help of frequentist and Bayesian statistics.

\begin{figure}[p]
\centering
\includegraphics[width = \linewidth]{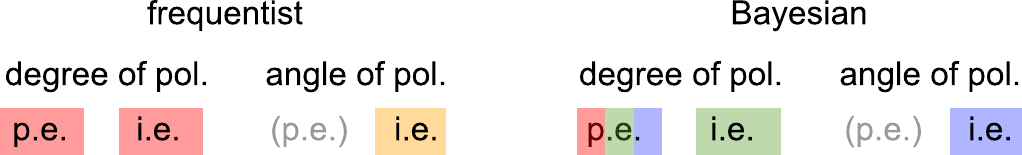}
\caption{Point and interval estimations (p.e. and i.e.) for the degree and angle of polarization result in four key values that can be addressed by a frequentist and a Bayesian approach. The existing works by 
\citet[][]{1985A&A...142..100S} 
(red), 
\citet[][]{1993A&A...274..968N} 
(orange), and 
\citet[][]{2006PASP..118.1340V} 
(green) are labeled with colored boxes. 
The contribution of this work is colored in blue. Point estimations for the angle of polarization are trivial and are not mentioned explicitly in literature, see \S\,\ref{sec:point_est_psi} and \S\,\ref{sec:point_est_psi_bay}.
}
\label{fig:freq_bay_family}
\end{figure}

\citet[][]{1950ratr.book.....C} rediscovered the Stokes parameters in 1950. Statistical considerations on the interpretation of polarimetric measurements are still an ongoing point of discussion. 
Figure 1 summarizes the main achievements in the field to which this work refers.
\citet[][]{1958AcA.....8..135S} 
emphasized that the observed degree of polarization is subject to biasing effects in a way that the observed value is preferentially greater than the true one. 
\citet[][]{1985A&A...142..100S} 
presented several point estimators to correct for this effect. Furthermore, they constructed confidence intervals to determine the reliability of the estimated degrees of polarization. These works are based on frequentist statistics.
\citet[][]{1993A&A...274..968N} 
constructed confidence intervals for the angle of polarization as a function of the true value of polarization.
These works rested on the assumption that the true value of polarization is known a priori; this is a fundamental concept in frequentist statistics.

However, in many situations with unknown true values, there is a need to construct point and interval estimations as functions of observational data.
Based on a Bayesian approach, \citet[][]{2006PASP..118.1340V} 
proposed a method to construct credibility intervals that, in comparison to the classical confidence intervals, show significant differences in the region of low signal-to-noise ratios. Using Bayesian analysis,
\citet[][]{2012A&A...538A..65Q} showed a complete summary of probability distributions for the degree and angle of polarization and possible priors suited to the experiment. The aim of this work is to extend the Bayesian approach to point estimations on the degree of polarization and to interval estimations for the angle of polarization. 

\S\,\ref{sec:statistics} summarizes the foundations of the frequentist and the Bayesian approach.
\S\,\ref{sec:stat} presents the underlying statistics that governs polarimetric measurements. 
Methods to construct point and interval estimations on the basis of frequentist statistics are reviewed in \S\,\ref{sec:freq} and recalculated with a Bayesian approach in \S\,\ref{sec:bay}. This section includes a new method to choose the best estimator for the degree of polarization and a new method to construct interval estimations on the polarization angle. Monte Carlo simulated credibility intervals for the degree and angle of polarization are presented in \S\,\ref{sec:sim}.
Results and conclusions follow in \S\,\ref{sec:res} and \S\,\ref{sec:con}. 

\section{Frequentist and Bayesian approach}
\label{sec:statistics}
The basis of frequentist statistics are probabilities of random events. Deductive reasoning leads to systems where unknown consequences can be studied for a known cause. In the example of polarimetric measurements, this means that the true degree and angle of polarization of radiation that enters a polarimeter are known and the different possible manifestations within the observed data are investigated. 

In contrast, Bayesian statistics extends the concept of probabilities to statements that become not only true and false, but more or less plausible. This reasoning allows one to deduce the plausibility of the cause on the basis of observed consequences. This is the usual case for all physical measurements: true values are estimated on the basis of observed data. In the case of polarimetric measurements, the true degree and angle of polarization are unknown parameters that shall be determined with the help of observed data. 

Thus, frequentist and Bayesian statistics address two different points of view. A direct comparison of both methods is not reasonable because both methods result in different statements. To find the appropriate method for a given problem, one has to answer the question of whether the true values or the observables are known or unknown.

\section{Statistics and Polarimetry}
\label{sec:stat}
\noindent

The normalized Stokes parameters $q$ and $u$ are appropriate variables to describe the linear polarization state of electromagnetic radiation (see \citet[][]{1983A&A...126..260C} for more details). The degree and angle of polarization, $P$ and $\varPsi$, can be expressed as
\begin{eqnarray}
	\label{eq:dop_stokes}
	P &=& \sqrt{q^2 + u^2} \hspace{8mm} \mathrm{and}\\
	\label{eq:psi_stokes}
	\varPsi &=& 0.5 \arctan{\big(u/q\big).}
\end{eqnarray}
Eq. (\ref{eq:dop_stokes}) clearly shows that random noise $\sigma$ in $q$ and $u$ results in a positively biased degree of polarization. $\sigma_\mathrm{q} = \sigma_\mathrm{u}$ is assumed. 
As the outcome of all type of polarimeters can be expressed in terms of Stokes parameters \citep[][]{2006PASP..118.1340V}, 
the following considerations are independent on the specific polarimeter type.

The applied terminology for a variable $X$ is the following: $x_0$ labels the true value of $X$, \,$\hat{x}$ is the estimated value, and $x$ is the observed value. The main variables are:\vspace{1mm}
   \[
      \begin{array}{lp{0.77\linewidth}}
         P              & degree of polarization \\
         \varPsi        & sky-angle of polarization  \\
         p              & signal-to-noise ratio of $p$ \, (Eq.(\ref{eq:p}))\\
         \sigma         & error in $q$ and $u$: $\sigma = \sigma_\mathrm{q} = \sigma_\mathrm{u}$\\
         \rho_{\Psi,\, (\mathrm{p})} & probability density of $\varPsi$ (or of $p$)\\
          \rho(X\,|\,A) & probability density of a variable set $X$ given the parameter set $A$ 
      \end{array}
   \]

   \begin{figure}[t]
  \centering
  \includegraphics[width = 1\linewidth]{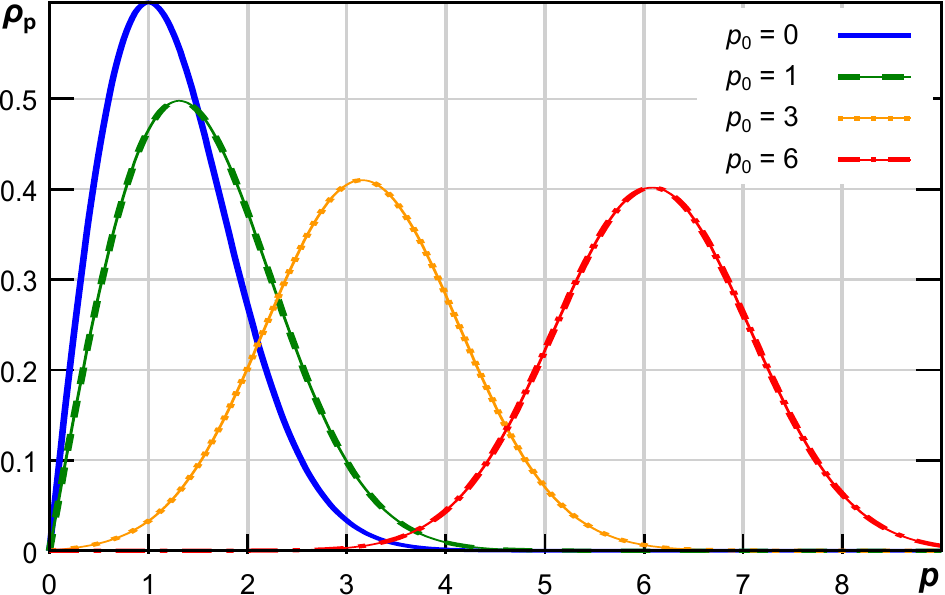}
  \caption{Probability density function $\rho_\mathrm{p}$ for different $p_0$, see also Eq.(\ref{eq:dp}). For large values of $p_0$, the distributions become Gaussian shaped.}
  \label{fig:dp}
  \end{figure}

\subsection{The signal-to-noise ratio of polarization}
\label{sec:snr_p}
The probability densities used in the next sections depend on different parameters that determine their shape. These parameters are the degree of polarization of the source $P$, 
and the uncertainties of the measurement $\sigma$. They can be combined to derive the signal-to-noise ratio
\begin{equation}
	\label{eq:p}
	p = \frac{P}{\sigma}.
\end{equation}
The value of $\sigma$ depends on the specific polarimeter type and observational conditions. 
See, for example, \citet[][]{2012SPIE.8443E..4NE} 
for a calculation of $\sigma$ for a counting based measurement with known background in the limit of low polarization.

With Eq. (\ref{eq:p}), the signal-to-noise ratio of any source observed with a specific polarimeter in a defined observation can be calculated. The signal-to-noise ratio $p$ is independent from specific experimental conditions and serves as appropriate basis for the following statistical treatment. 

  \begin{figure}[t]
  \centering
  \includegraphics[width = 0.99\linewidth]{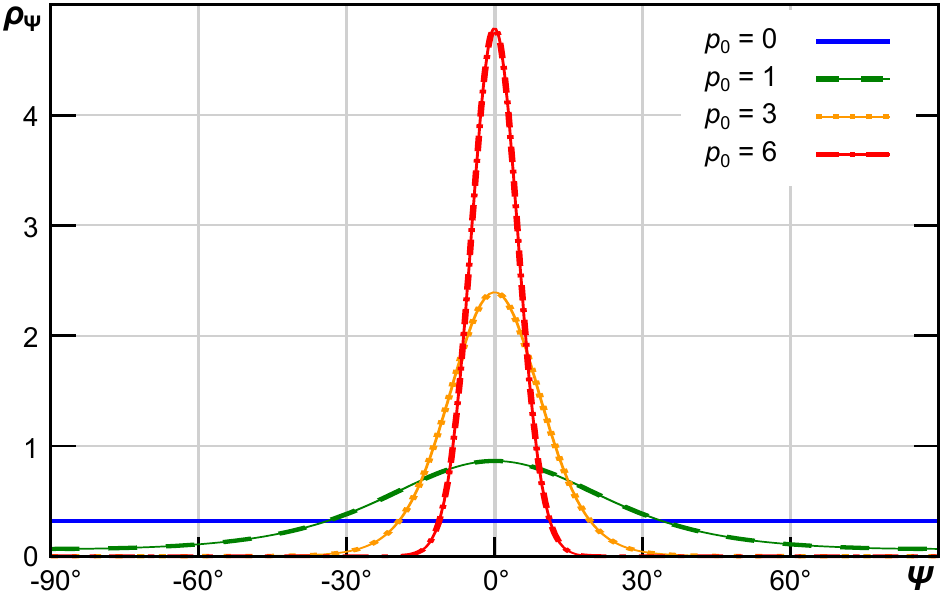}
  \caption{Probability density function $\rho_\Psi$ for different $p_0$, see also Eq.(\ref{eq:dPsi}). The distributions are non-Gaussian, but symmetrical around $\varPsi_0$ (here: $\varPsi_0 = 0^\circ$).}
  \label{fig:dPsi}
  \end{figure}

\subsection{The probability density function \texorpdfstring{$\rho$}{rho}}
The detection principle of a polarimeter is to measure a sinusoidal variation of the signal\footnote{The following calculations are based on the ideal case that observed polarimetric data are distributed continuously. Effects resulting from discontinuities, such as data binning, are not considered.} intensity as a function of the azimuthal angle. Based on that idea, \citet[][]{2012SPIE.8443E..4NE} derived, in the case of low polarizations $(P^2 \ll 1)$, the bivariate probability density function $\rho$ which represents the probability of observing a signal-to-noise ratio $p$ within the interval $[p, p+\mathrm{d}p]$ and a polarization angle $\varPsi$ within $[\varPsi, \varPsi+ \mathrm{d}\varPsi]$ while the true values are $p_0$ and $\varPsi_0$:
\begin{eqnarray}
	\label{eq:pdf}
	&&\hspace{-6.6mm}\rho(p,\varPsi\,|\,p_0,\varPsi_0) \, \mathrm{d}p \, \mathrm{d}\varPsi = \\
	\nonumber && \frac{p}{\pi} \exp{\left(-\frac{(p - p_0)^2 + pp_0\big(2 \sin(\varPsi-\varPsi_0)\big)^2}{2}\right)} \, \mathrm{d}p \, \mathrm{d}\varPsi.
\end{eqnarray}
Equivalent results\footnote{This work is always related to the sky-angle and not the angle in the q-u-plane. Because $-\piup/2 < \Psi \le \piup/2$ holds for Eq. (\ref{eq:dp}) a factor of $1/2$ is missing compared to other works that consider $-\piup < \Psi \le \piup$.} obtained under the assumption of Gaussian distributed Stokes parameters $q$ and $u$, but without any limitations on $p$, are presented by \citet[][]{2012A&A...538A..65Q}.
Integrating Eq. (\ref{eq:pdf}), with respect to $p$, yields the Rice distribution \citep[][]{1945BSTJ...24...46R} -- the univariate probability density function for the degree of polarization\footnote{Statistical considerations on the degree of polarization (see \S\,\ref{sec:dp} and \S\,\ref{sec:dp_bay}) can be generalized to any fields in which a quantity is estimated from a quadrature sum of other quantities.} $\rho_\mathrm{p}$ 
\citep[][]{2012SPIE.8443E..4NE} 
-- while integrating Eq. (\ref{eq:pdf}), with respect to $\varPsi$, results in the univariate probability density function for the angle of polarization 
$\rho_\Psi$ \citep[][]{1993A&A...274..968N}: 
  \begin{eqnarray}
  \label{eq:dp}
        \rho_\mathrm{p}(p\,|\,p_0) \hspace{-2.5mm}&=&\hspace{-2.9mm} p \cdot \exp{\left( -\frac{p^2 + p_0^2}{2}\right) \cdot \mathrm{I}_{1,\,0}(p p_0)}\hspace{2mm}\\
  \label{eq:dPsi}
  \rho_{\Psi}(\varPsi\,|\,p_0,\varPsi_0) \hspace{-2.5mm}&=&\hspace{-2.9mm} \frac{1}{\sqrt{\pi}}\!\left(\!\frac{1}{\sqrt{\pi}}\!+\!\eta \, \mathrm{e}^{\eta^2} \big(1\!+\!\mathrm{erf}(\eta)\!\big)\!\right) \mathrm{e}^{-p_0^2/2} \hspace{2mm}
  \end{eqnarray}
where $\mathrm{I}_{1,\,0}$ is the modified Bessel function of first kind and zeroth order, erf is the Gauss error function, and $\eta =~p_0 /\! \sqrt{2} \cdot \cos\big(2(\varPsi - \varPsi_0)\big)$. The probability density functions are plotted in Fig. \ref{fig:dp} and \ref{fig:dPsi}.
  

\begin{figure}[t]
\centering
\includegraphics[width = 0.978\linewidth]{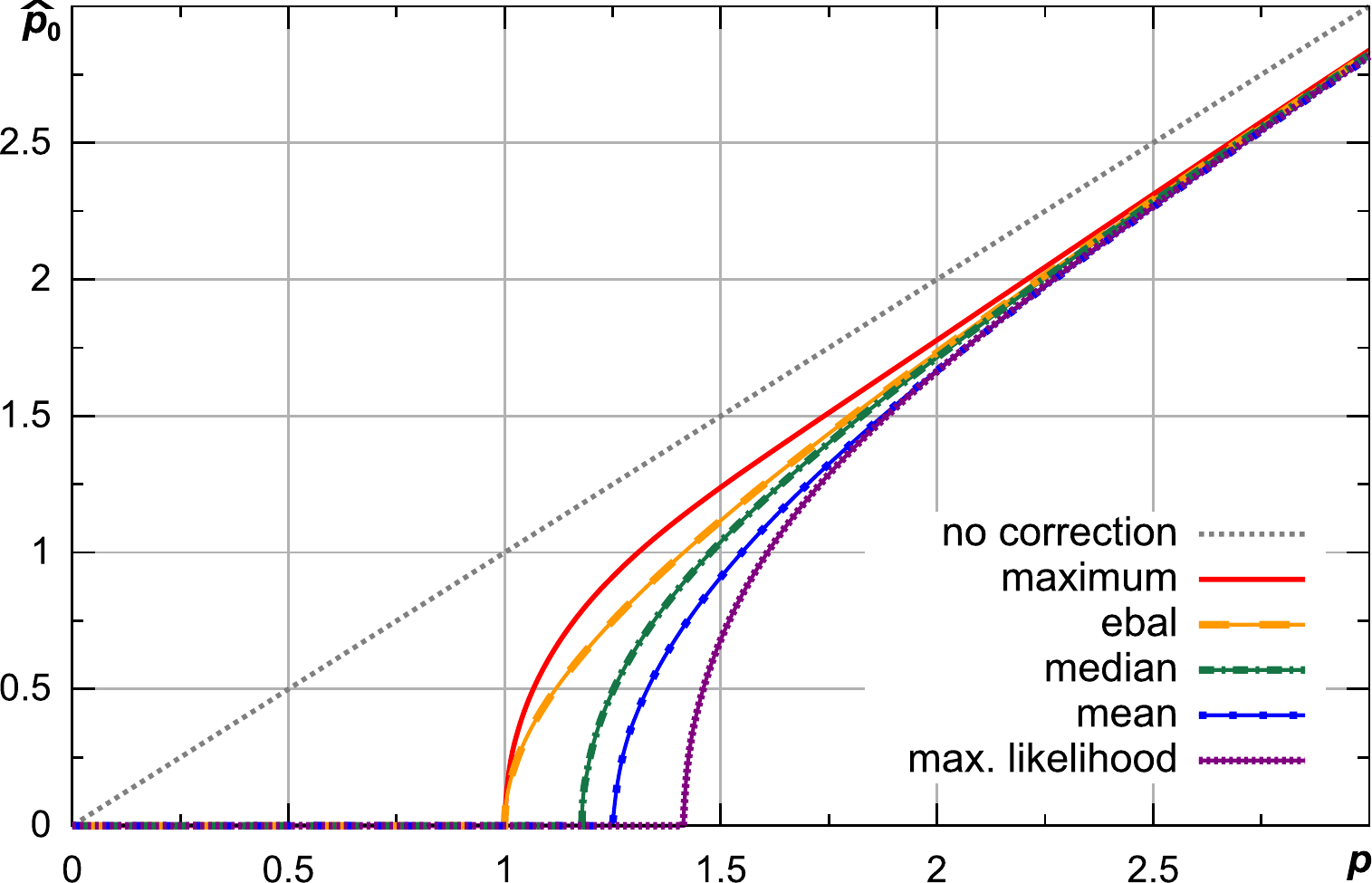}
\caption{Plots of the maximum, median, mean, and maximum likelihood estimator presented in \citet[][]{1985A&A...142..100S} 
in comparison with the ebal-estimator (Eq. (\ref{eq:high_approx})).
}
\label{fig:p_estimation}
\end{figure}

\section{Point and interval estimation on the basis of frequentist statistics}
\label{sec:freq}

\subsection{Degree of polarization}
\label{sec:dp}
The $p$-distribution $\rho_\mathrm{p}$, in Fig. \ref{fig:dp}, shows that even unpolarized light $(p_0 = 0)$ will yield an observed polarization $p > 0$. \citet[][]{1985A&A...142..100S} 
showed several methods to correct for this biasing effect. They also presented a suggestion on how to construct confidence intervals around an estimated polarization. For completeness, we briefly review their results.

\subsubsection{Point estimation}
\begin{figure}[t]
\centering
\includegraphics[width = 1\linewidth]{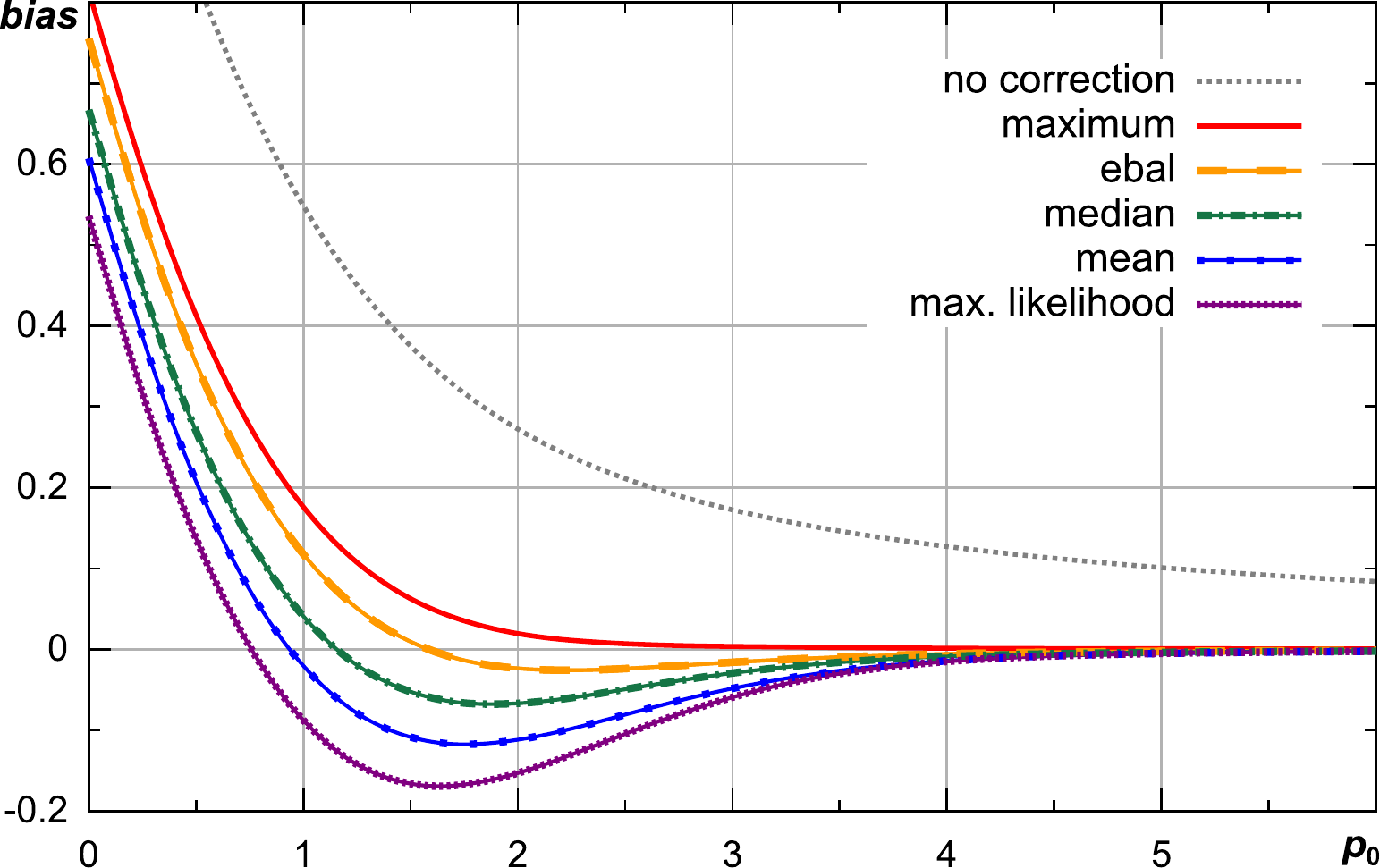}
\caption{Bias of the estimators presented in \citet[][]{1985A&A...142..100S} 
in comparison with the ebal-estimator (Eq. (\ref{eq:high_approx})). The max. likelihood estimator works best for low signal-to-noise ratios, whereas the maximum estimator works best for high values of $p_0$. }
\label{fig:bias}
\end{figure}

An estimator $\hat{p}_0$ can be constructed in different ways. \citet[][]{1985A&A...142..100S} 
presented estimators based on the maximum, the median, and the mean of the $\rho_\mathrm{p}(p\,|\,p_0)$ distribution. They also proposed an estimator based on the maximum of the corresponding likelihood function\footnote{Cf. Eq. (\ref{eq:post_distr1}) and Eq. (\ref{eq:post_distr2}) and footnote \ref{fn:ss85}.} 
$\rho_{\mathrm{p}_0}(p_0\,|\,p)$. In addition to this, we present a very simple estimator \textit{ebal}\footnote{\textbf{e}stimator \textbf{b}ased on \textbf{a}pproximated max. \textbf{l}ikelihood.} that uses an approximate behavior of the maximum likelihood estimator for high signal-to-noise ratios \citep[][]{2006PASP..118.1340V} 
in combination with a cut-off at $p = 1$:
\begin{eqnarray}
	\label{eq:high_approx}
  \notag
	\hat{p}_{0,\,\textnormal{ebal}}(p) &=& 0  \hspace{27.7mm} p \le 1\\
	&=& \sqrt{p^2 - 1}  \hspace{17mm} p > 1.
\end{eqnarray} 
Polynomial fits for the other estimators are listed in \citet{1991A&A...246..280S}. 
Fig. \ref{fig:p_estimation} shows the graphs of all estimators. The estimated difference between the estimation value and the true value, the so-called $bias$, can be used to decide which estimator works best:
   \begin{equation}
   \label{eq:bias}
	   bias(p_0) = \int\limits_0^{\infty}\! \rho_\mathrm{p}(p\,|\,p_0) \cdot     
	   \hat{p}_0(p) \,\,\mathrm{d} p \,-\, p_0.
   \end{equation}
The best estimator is the one with the smallest bias.
Fig. \ref{fig:bias} shows that all estimators work best in different regions of $p_0$. \citet[][]{1985A&A...142..100S} 
concluded that the maximum likelihood estimator should be used for low signal-to-noise ratios and the maximum estimator for high signal-to-noise ratios $p_0$. This reasoning is not practically applicable because the estimators are used to obtain values for $p_0$, so $p_0$ is unknown at the moment the best estimator must be chosen. See \S\,\ref{sec:dp_point_bay} for a solution to this problem.
Beside this problem, \citet[][]{1985A&A...142..100S} did not consider the promising ebal-estimator, which has a bias smaller than 2.4\,\% for signal-to-noise ratios $p_0 > 1.4$, and also has a simple analytical description.

\subsubsection{Confidence intervals}
When the best estimator $\hat{p}_0$ for an observed polarization $p$ is known, the confidence interval makes a statement about the reliability of this estimation.
The definition of a confidence interval $\Delta p_0 = \big[\underline{p_0},\, \overline{p_0}\big]$, with confidence level $C$, is related to a set of repeated measurements of an observable $p$ for a true, but unknown, parameter $p_0$. The probability that the constructed $(\Rightarrow)$ confidence intervals, corresponding to the observables $p$, contain the true value $p_0$ is $C$:
\begin{equation}
	\label{eq:conf_int}
	\forall \, (p\,\Rightarrow \Delta p_0): \; \mathrm{Prob}\left(p_0 \in \Delta p_0\right) = C \, \mathrm{,fixed }\; p_0.
\end{equation}
In analogy with the Gaussian distribution, confidence intervals are labeled as $\sigma_1$, $\sigma_2$, and $\sigma_3$ for $C = 68.3\,\%$, $C = 95.5\,\%$, and $C = 99.7\,\%$. It should be noted that $\sigma_n \ne n \cdot \sigma_1$ as in the Gaussian case.

Confidence intervals make a statistical statement about the \emph{intervals} containing the true value, but not about the \emph{true value} being within a specific interval.
The construction of a confidence interval is not a trivial calculation because it is related to the true polarization $p_0$, but the distribution $\rho_\mathrm{p}(p\,|\,p_0)$ (Eq. (\ref{eq:dp})) is a function of $p$.
\citet[][]{1985A&A...142..100S} 
constructed intervals $\Delta p = \big[\underline{p},\,\overline{p}\big]$ by integrating $\rho_\mathrm{p}$ numerically, so that
\begin{equation}
	\label{eq:int_simmons}
	\int\limits_{\underline{p}}^{\overline{p}} \rho_\mathrm{p}(p\,|\,p_0)\,\mathrm{d} p = C \hspace{7mm} \bigg| \quad \; \big[\underline{p},\, \overline{p}\big] \textnormal{ minimal.}
\end{equation}
\begin{figure}[t]
\centering
\includegraphics[width = \linewidth]{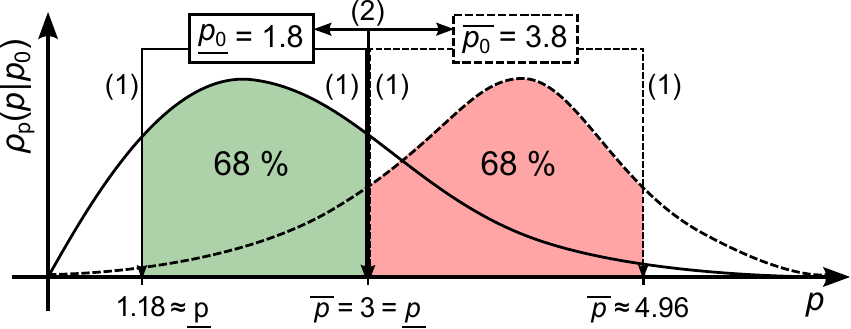}
\caption{Illustration of the $p_0$-$p$ projection (\citet[][]{1985A&A...142..100S}) for a $\sigma_1$ confidence interval in the case $p=3$. If the true polarization is $p_0 = 1.8$ then 68 \% of all observed polarizations $p$ lie within $\Delta p = [1.18, 3]$. The corresponding interval for $p_0 = 3.8$ is $\Delta p~=~[3, 4.96]$. These calculations are based on Eq. (\ref{eq:int_simmons}). The $\sigma_1$ confidence interval for $p = 3$ is therefore $\Delta p_0 = [1.8, 3.8]$. See also Fig. \ref{fig:CI}.}
\label{fig:ss_expl}
\end{figure}

\noindent
The additional demand that $\Delta p$ should be as narrow as possible is due to the fact that Eq. (\ref{eq:int_simmons}) is valid for a infinite number of intervals $\Delta p$.
This demand, in combination with the asymmetric distribution $\rho_\mathrm{p}$, results in intervals $\Delta p$ that are not symmetric with respect to the maximum of $\rho_\mathrm{p}$, as well.

\begin{figure}[t]
\centering
\includegraphics[width = 1\linewidth]{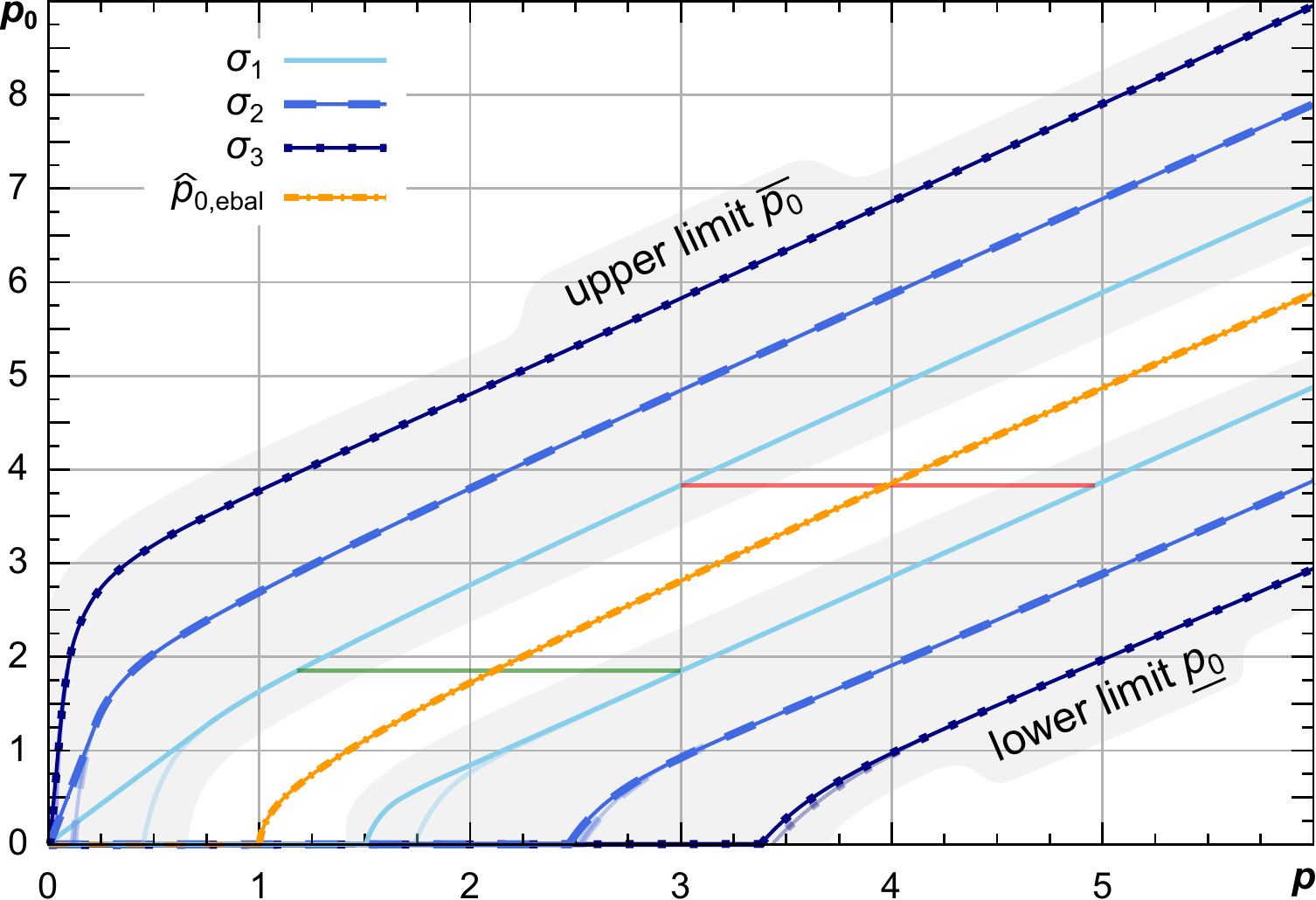}
\caption{Confidence intervals $\big[\underline{p_0},\, \overline{p_0}\big]$ in terms of $\sigma_1$, $\sigma_2$, and $\sigma_3$ as function of the observed polarization $p$. The x/y-coordinates are interchanged compared to the similar plots shown by \citet[][]{1985A&A...142..100S}
. The ebal estimator (orange graph in the center) shows that estimated values are in general not centered within the confidence interval. This becomes more prominent for small values of $p$ and large confidence levels. 
The original construction method (Eq.~(\ref{eq:int_simmons})) is indicated in pale colors in contrast to the tangential progression (bold colors). The interval examples of Fig. \ref{fig:ss_expl} are plotted with consistent colors for illustrative reasons.
Example: for an observed polarization of $p = 3 $ the corresponding $\sigma_1$-confidence interval is approximately $\Delta p_0 \approx  [1.8, 3.8]$ while the estimated polarization is $\hat{p}_0 \approx 2.83$ by use of the ebal estimator. See Table \ref{tab:fit_f} for parametric fitting results for $\underline{p_0}$ and $\overline{p_0}$.
}
\label{fig:CI}
\end{figure}

Now, instead of an interval $\Delta p$ obtained for a given $p_0$, an interval $\Delta p_0$ for a given $p$ is needed. This projection 
is done in two steps:
\begin{enumerate}
\item For all $p_0$ values corresponding $\underline{p}$ and $\overline{p}$ can be calculated with Eq. (\ref{eq:int_simmons}).
\item For a specific $p$ value, only two $p_0$-$\underline{p}$-$\overline{p}$ triplets are of interest: the one triplet with its upper limit $\overline{p}$ equal to $p$ defines the lower limit of $p_0$ ($\underline{p_{0}}$); and the other triplet with its lower limit $\underline{p}$ equal to $p$ defines the upper limit of $p_0$ ($\overline{p_0}$), see Fig.~\ref{fig:ss_expl}.
\end{enumerate}
The following example may illustrate the situation for $\sigma_1$ confidence intervals (see Fig. \ref{fig:CI} in pale colors): for $p = 3$, $\Delta p_0 \approx [1.8, 3.8]$ is well defined. For $p = 1.25$, the $p_0$-projection (Fig. \ref{fig:ss_expl}) is no longer applicable because the lower limit is not zero, but undefined. Nevertheless, the confidence limit $\Delta p_0 \approx [0, 1.9]$ is in accordance with Eq.~(\ref{eq:conf_int}), if it is assumed that $\Delta p_0$ does not exist for $p \lesssim 0.4$ (the intersection point of the upper limit curve with the $p$-axis).

To avoid the case of non-existing confidence limits, \citet[][]{1985A&A...142..100S} 
departed from the term of narrowest $\Delta p$ for small values of $p_0$. They proposed a tangential progression of the upper limit $\overline{p_0}$ to zero for $p \rightarrow 0$ and calculated corresponding new values of $\underline{p_0}$, so that the definition of Eq.~(\ref{eq:conf_int}) still holds. This way, confidence intervals can be set for all $p > 0$. 
Fig.~\ref{fig:CI} shows $\sigma_1$, $\sigma_2$, and $\sigma_3$ confidence intervals with and without the tangential progression. See Tab.\ref{tab:fit_f} for parametric fitting results.

Another example shows the interpretation of confidence intervals: for $p=0.25$ the $\sigma_1$ confidence interval (by tangential construction) is $\Delta p_0 \approx [0, 0.4]$. This does not mean that the observed radiation is measured very precisely, but that a set of repeated measurements will result in  different -- most likely higher -- values of $p$ and new $\Delta p$ that contain the true value $p_0$ in about 68\,\% of all cases.

\subsection{Angle of polarization}
\subsubsection{Point estimation}
\label{sec:point_est_psi}
Since there is no biasing effect in the angle of polarization (see Fig. \ref{fig:dPsi}), the trivial estimator $\hat{\varPsi}_0(\varPsi) = \varPsi$ is sufficient.
\subsubsection{Confidence intervals}
\begin{figure}[t]
\centering
\includegraphics[width = 1\linewidth]{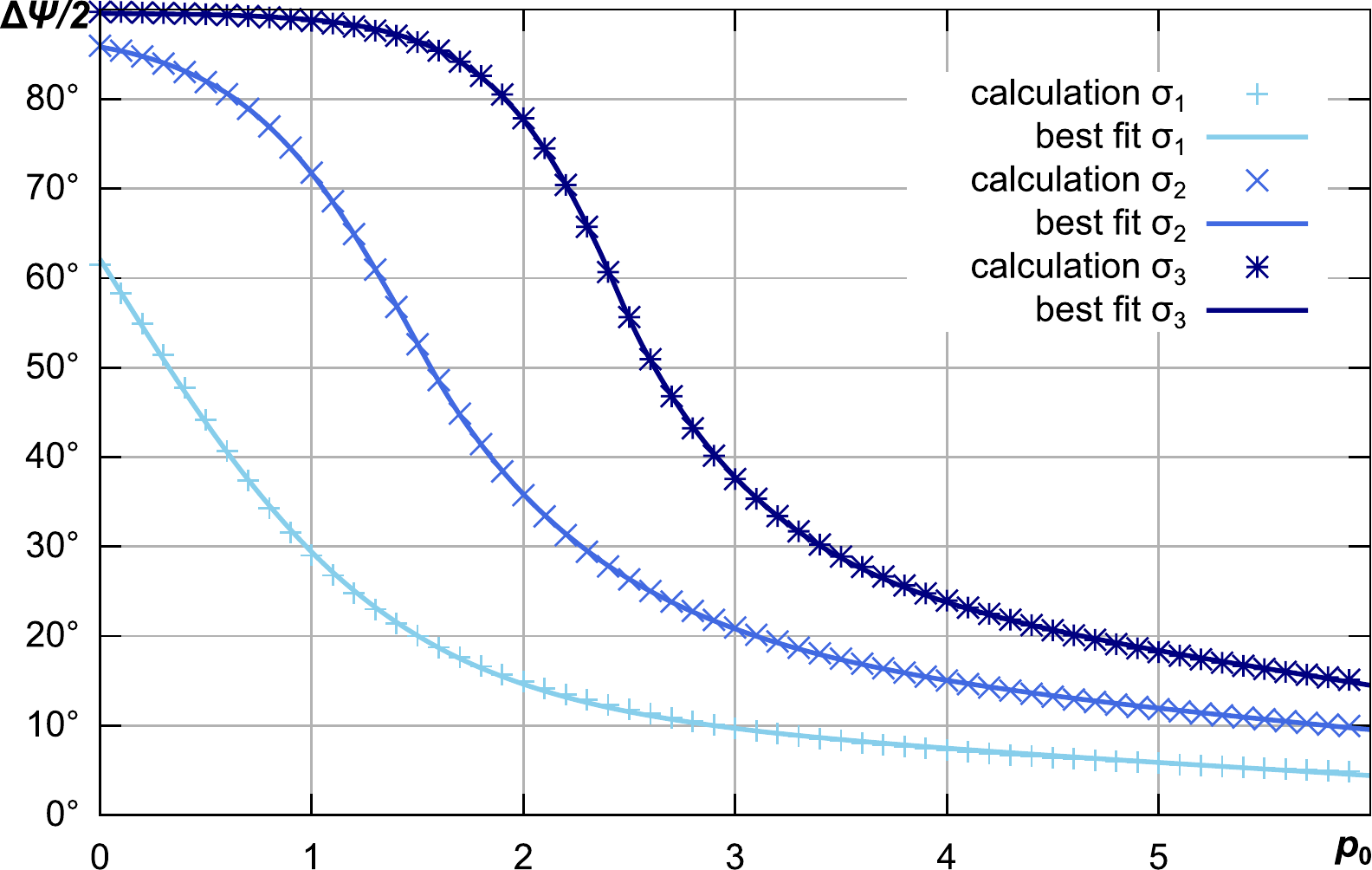}
\caption{Uncertainties in $\varPsi$ in terms of $\sigma_1$, $\sigma_2$, and $\sigma_3$ confidence intervals as function of $p_0$, (Eq. (\ref{eq:Psi_sigma})). The plots were presented first by \citet[][]{1993A&A...274..968N} 
(be aware of the wrong labeling in their Fig. 2a). The confidence intervals are $\Delta\varPsi = \big[\varPsi_0 - \Delta\varPsi/2,\, \varPsi_0 + \Delta\varPsi/2\big]$.}
\label{fig:Psi_sigma}
\end{figure}

\citet[][]{1993A&A...274..968N} 
presented plots for $\Delta \varPsi$ equal to  $\sigma_1$, $\sigma_2$, and $\sigma_3$ confidence intervals $\big[\varPsi_0 - \Delta\varPsi/2,\, \varPsi_0 + \Delta\varPsi/2\big]$ by numerically integrating $\rho_\Psi$ (see Fig. \ref{fig:Psi_sigma}):

\begin{equation}
	\label{eq:Psi_sigma}
	\int_{\varPsi_0-\Delta \varPsi/2}^{\varPsi_0+\Delta \varPsi/2} \rho_\Psi(\varPsi\,|\,p_0, \varPsi_0) \, \mathrm{d} \varPsi = C.
\end{equation}
\vspace{1.5mm}

\noindent
The basic idea in Eq. (\ref{eq:Psi_sigma}) is the same as in Eq.~(\ref{eq:int_simmons}), with the only difference being that the symmetry of $\rho_\Psi$ around $\varPsi_0$ makes the narrowest confidence intervals symmetric around $\varPsi_0$, as well.
The dependence of $\rho_\Psi$ on $p_0$ is difficult because $p_0$ is unknown from an observational point of view.

\noindent
Even though the plots of Fig. \ref{fig:Psi_sigma} are a function of $p_0$ and therefore not directly usable for observational data, 
it should be noted that the uncertainty of the polarization angle is a strictly monotonically decreasing function of $p_0$.
As Eq. (\ref{eq:Psi_sigma}) is independent of $p$, we see no way to construct confidence intervals for $\varPsi_0$ as a function of~$p$. See hereto the discussion in \S\,\ref{sec:angle_cred_int}.

\section{Point and interval estimation on the basis of Bayesian statistics}
\label{sec:bay}
So far, the frequentist approach -- taking the true polarization $p_0$ as a fixed parameter -- leads to the strange situation that the unknown true polarization $p_0$ must be known, or at least estimated, to select the best estimator $\hat{p}_0$. Furthermore, the frequentist approach with it's $p-p_0$ projection, results in a difficult construction of confidence intervals $\Delta p_0$ at low signal-to-noise ratios $p$ and it seems impossible to construct usable confidence intervals $\Delta \varPsi$ as a function of~$p$. 

All of these problems disappear with a Bayesian approach because now, the true polarization $p_0$ can be treated as a stochastic variable.
\begin{figure}[t]
\centering
\includegraphics[width = \linewidth]{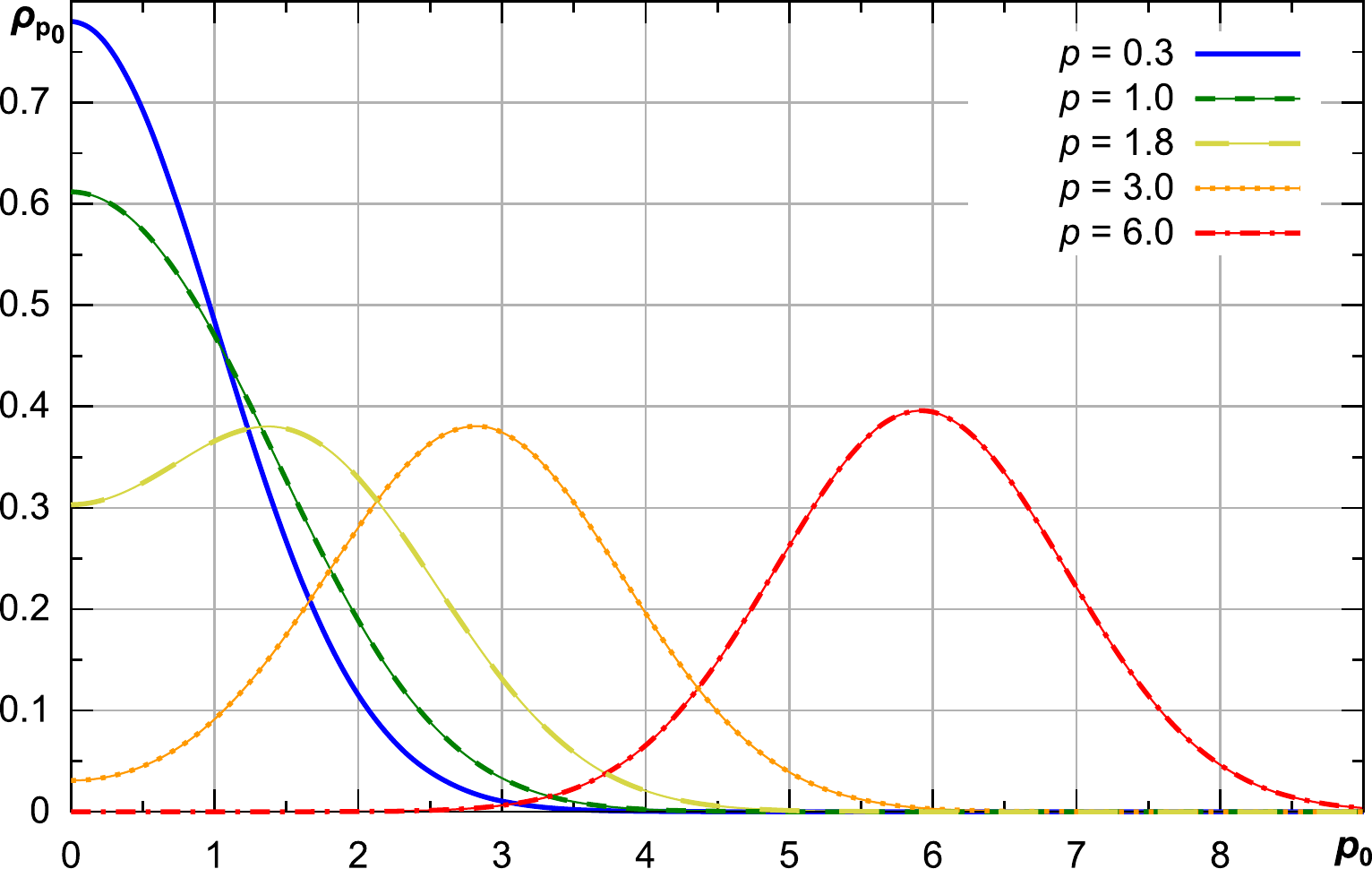}
\caption{The posterior density distribution $\rho_{\mathrm{p}_0}(p_0\,|\,p)$ represents the probability density that an observed polarization $p$ results from a true polarization $p_0$. These distributions (Eq. (\ref{eq:post_distr1})) were first calculated by \citet[][]{2006PASP..118.1340V}
.}
\label{fig:pdf_p0}
\end{figure}
In general, the posterior density $\rho_{\mathrm{p}_0}(p_0\,|\,p)$ can be computed with the likelihood $\rho_\mathrm{p}(p\,|\,p_0)$ and the prior density $\rho(p_0)$ with Bayes theorem:
\begin{equation}
	\rho_{\mathrm{p}_0}(p_0\,|\,p) = \frac{\rho(p_0) \cdot \rho_\mathrm{p}(p\,|\,p_0)}{\int_{0}^{\infty} \rho(p_0) \cdot \rho_\mathrm{p}(p\,|\,p_0) \,\, \mathrm{d} p_0}.
\end{equation}
In the following, we are considering a non-informative polar prior density:  $\rho(p_0)\!\!~=~\!\!\mathrm{const.}$\footnote{Technically speaking, this statement is critical because there is no uniform distribution living on the non-negative half-line R$_+$ that can be normalized. This mathematical problem can be overcome by considering a maximal possible true degree of polarization $p_{0,\mathrm{ max}}$, so that \mbox {$\rho= p_{0,\mathrm{ max}}^{-1}$}. Finally: $p_{0,\mathrm{ max}} \rightarrow \infty$.}.
\citet[][]{2012A&A...538A..65Q} studied the impact of non-informative prior densities, in general, and the difference of Jeffrey's prior, which is uniform in the Stokes parameters $q_0$ and $u_0$ to the uniform polar prior, which is uniform in $p_0$. With reference to this work, it shall only be mentioned that Jeffrey's prior overstates large values of $p_0$, so that the polar prior seems to be the best choice if any information from the source is missing:
\begin{eqnarray}
		\label{eq:post_distr1}
		\rho_{\mathrm{p}_0}(p_0\,|\,p) &=& \frac{1}{N} \cdot \rho_\mathrm{p}(p\,|\,p_0) \\
		\label{eq:post_distr2}
		 \textnormal{with } \hspace{5mm}N &=& \int\limits_{0}^{\infty} \!\rho_\mathrm{p}(p\,|\,p_0) \, \mathrm{d} p_0.
\end{eqnarray}
This is equivalent to a simple interchanging of parameter $p_0$ with variable $p$ in Eq.(\ref{eq:dp}), combined with a subsequent normalization\footnote{\label{fn:ss85}The normalization is unimportant for the calculation of the maximum likelihood estimator so that \citet[][]{1985A&A...142..100S} 
simply set $\rho_\mathrm{p_0} = \rho_\mathrm{p}$ for their calculations. 
}$N$  \citep[][]{2006PASP..118.1340V}
. Fig.~\ref{fig:pdf_p0} shows a sample of posterior density distributions of $\rho_{\mathrm{p}_0}$.

\subsection{Degree of polarization}
\label{sec:dp_bay}
\subsubsection{Point estimation}
\label{sec:dp_point_bay}
The probability density $\rho_{\mathrm{p}_0}$ can be used to transform any quantity that is a function of $p_0$ to an estimation value as a function of $p$. Using the already calculated $bias(p_0)$ of Eq. (\ref{eq:bias}) leads to
\begin{equation}
  \label{eq:proj}
	bias(p) = \int\limits_0^{\infty} bias(p_0) \cdot \rho_{\mathrm{p}_0}(p_0\,|\,p) \,\mathrm{d}p_0.
\end{equation}

\begin{figure}[t]
\centering
\includegraphics[width = 1\linewidth]{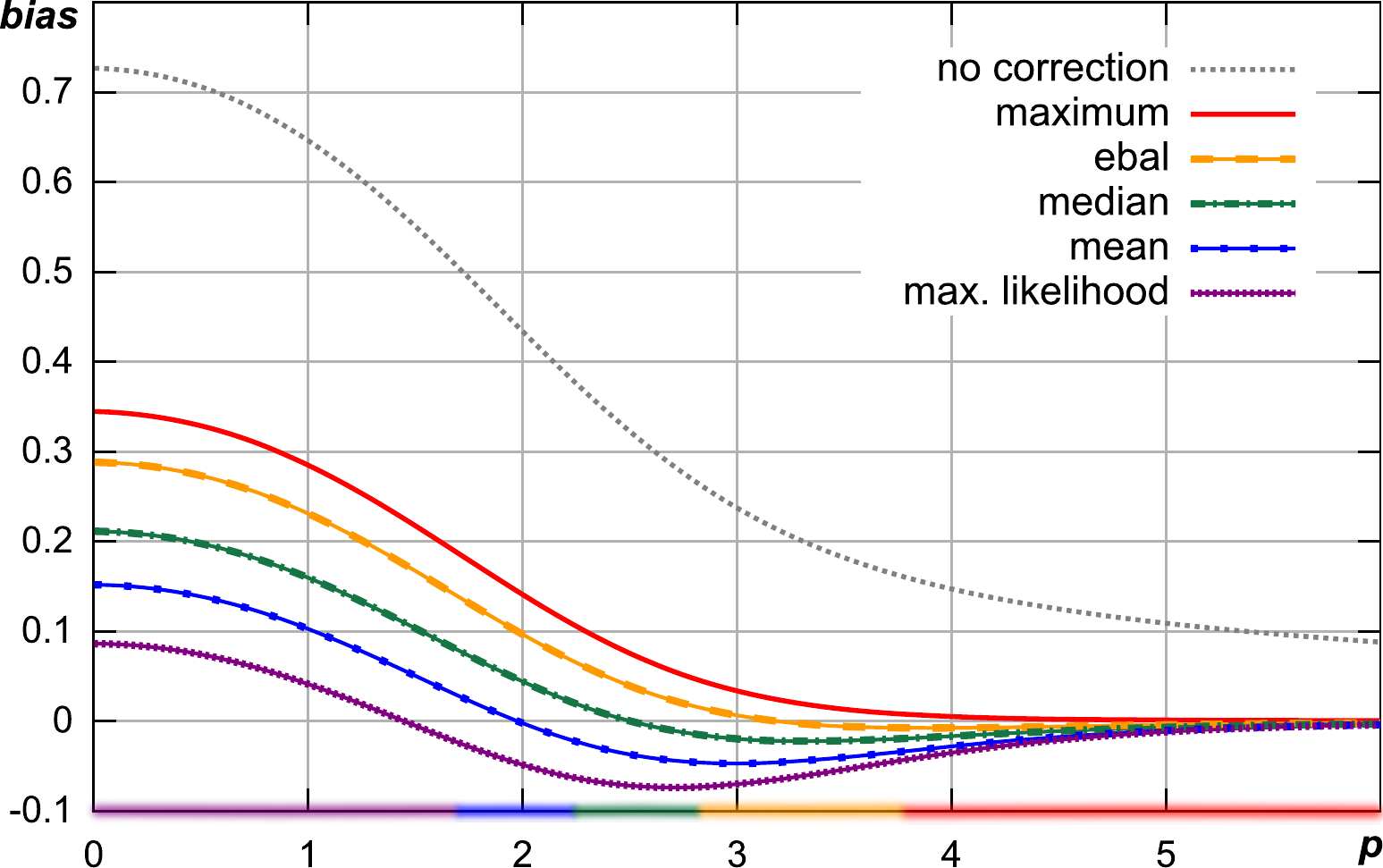}
\caption{Bayesian $bias$ as function of $p$. The maximum, median, mean, and maximum likelihood estimator are constructed like in \citet[][]{1985A&A...142..100S} 
The ebal-estimator is defined in Eq. (\ref{eq:high_approx}). The regions of best performance are indicated as color bars on the $p$-axis. See \S\,\ref{sec:degofpol} for listed values.}
\label{fig:offset_p}
\end{figure}

Each value of $bias(p_0)$ is weighted with the probability that this $p_0$ value results from an observed value $p$. Fig. 
The estimators in Fig. \ref{fig:offset_p} show a similar behavior to those presented in Fig. \ref{fig:bias},
but now, the regions of the best performance are in terms of $p$. This allows the best estimator to be chosen directly for the first time on the basis of the observed data, cf. \S\,\ref{sec:degofpol}.

\subsubsection{Credibility intervals}
\begin{figure}[t]
\centering
\includegraphics[width = 1\linewidth]{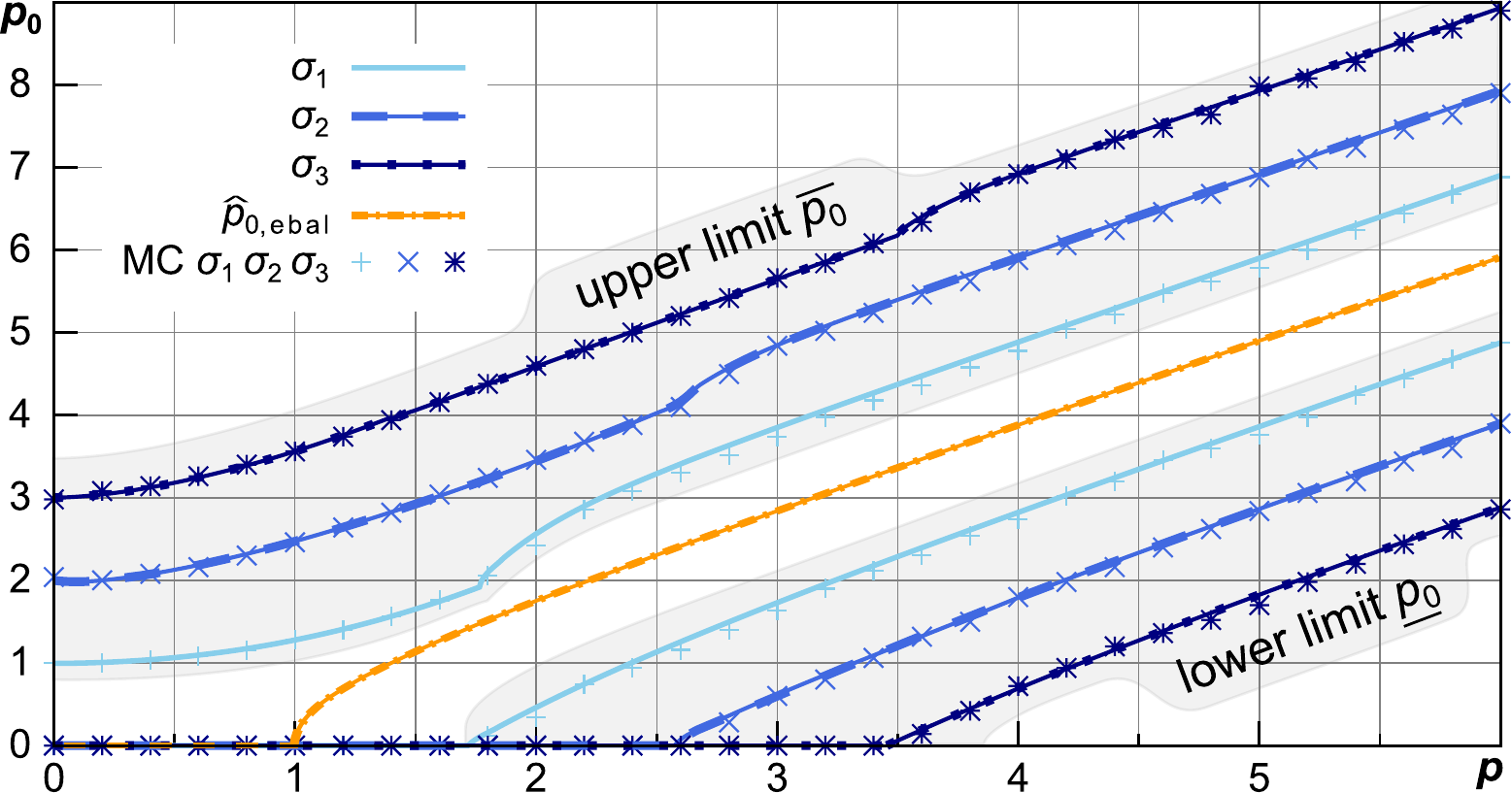}
\caption{$\sigma_1$, $\sigma_2$, and $\sigma_3$ credibility intervals $\big[\underline{p_0},\, \overline{p_0}\big]$ as function of $p$.
Example: for $p= 3$ the best performance is expected with the ebal estimator (see Fig.\ref{fig:offset_p}): $\hat{p}_{0,\, \mathrm{ebal}}(3) \approx 2.83$. The corresponding $\sigma_1$-interval is approximately $\Delta p_0 \approx  [1.7, 3.8]$.
The results of the Monte Carlo (MC) simulation of \S\,\ref{sec:sim} are plotted with points. See Table \ref{tab:fit_f_bay} for parametric fitting results for $\underline{p_0}$ and $\overline{p_0}$.}
\label{fig:CI_bay}
\end{figure}

Analogous to confidence intervals in frequentist statistics, credibility intervals in Bayesian statistics make a statement on the reliability of an estimated value. Different from confidence intervals, credibility intervals give the probability that the true value is within a specific interval. For all possible values of $p_0$ that can cause ($\twoheadrightarrow$) the observed $p$, the credibility interval includes those $p_0$ that cause $p$ in a fraction $C$ of all cases:
\begin{equation}
	\label{eq:cred_def}
	\forall \, p_0: \, \mathrm{Prob}\left(p_0 \in \big[\underline{p_0},\, \overline{p_0}\big] \twoheadrightarrow p\right) = C \;\,\, \mathrm{,fixed}\; p.
\end{equation}
Integrating the density distribution $\rho_{\mathrm{p}_0}$ over $p_0$ leads directly to credibility intervals \citep[][]{2006PASP..118.1340V}: 
\begin{equation}
	\int\limits_{\underline{p_0}}^{\overline{p_0}} \rho_{\mathrm{p}_0}(p_0\,|\,p)\,\mathrm{d} p_0 = C \hspace{4mm} \bigg| \quad \big[\underline{p_0},\, \overline{p_0}\big] \textnormal{ minimal.}
\end{equation}
The resulting credibility limits are shown in Fig. \ref{fig:CI_bay} for $\sigma_1$, $\sigma_2$, and $\sigma_3$ credibility intervals. See Table \ref{tab:fit_f_bay} for parametric fitting results. 

\subsection{Angle of polarization}
\subsubsection{Point estimation}
\label{sec:point_est_psi_bay}
As in \S\,\ref{sec:point_est_psi}, the trivial estimator $\hat{\varPsi}_0(\varPsi)~=~\varPsi$ is sufficient for point estimations on the angle of polarization.

\subsubsection{Credibility intervals}
\label{sec:angle_cred_int}
\begin{figure}[t]
\centering
\includegraphics[width = \linewidth]{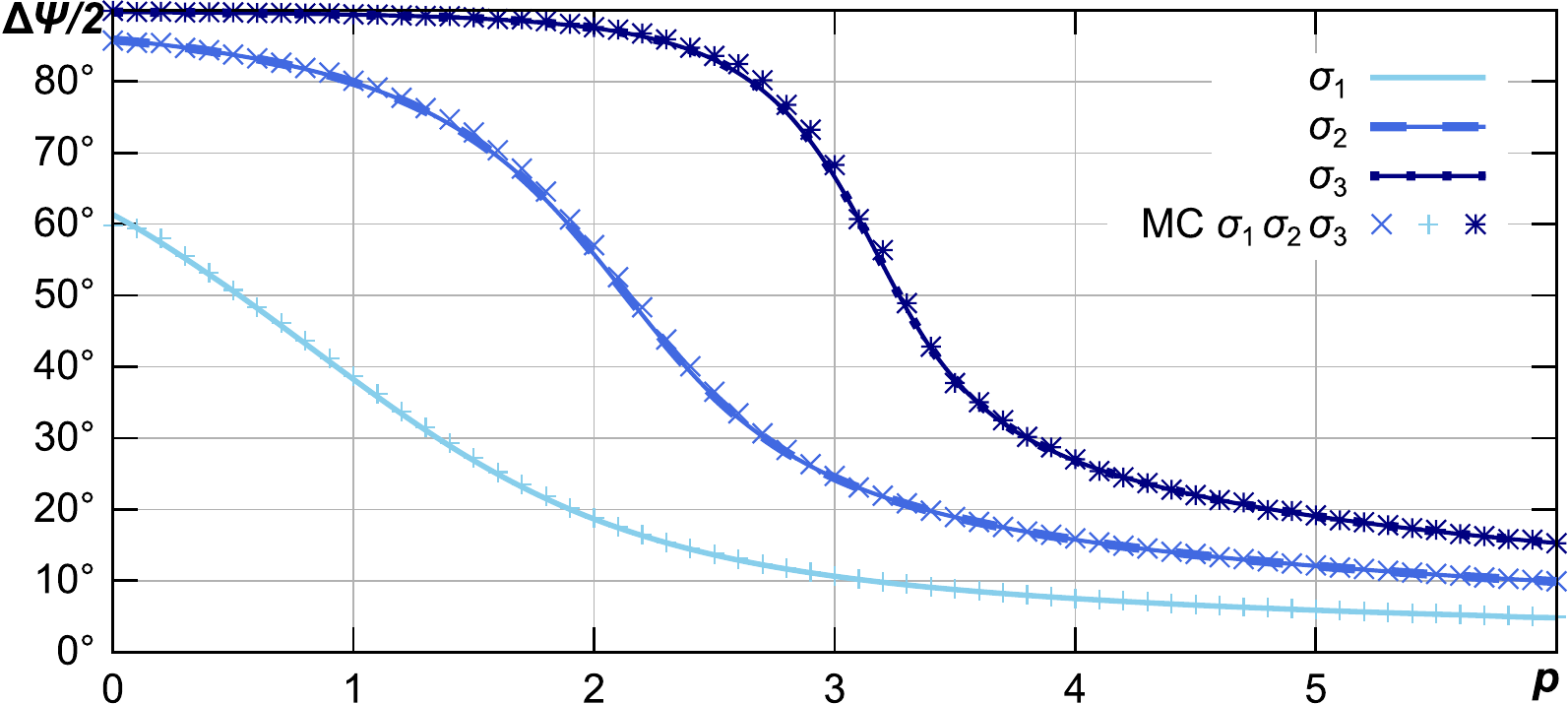}
\caption{$\sigma_1$, $\sigma_2$, and $\sigma_3$ credibility intervals $\big[\varPsi_0 - \Delta\varPsi/2,\, \varPsi_0 + \Delta\varPsi/2\big]$. The results of the Monte Carlo (MC) simulation of Section \ref{sec:sim} are plotted with points.}
\label{fig:Psi_sigma_cor}
\end{figure}

Before proposing our method, we want to list the difficulties we recognized in constructing interval estimations $\Delta \varPsi(p)$ and clarify why it is incorrect to proceed in the following way:
\begin{itemize}
\item Using the best estimator $\hat{p}_0$ as a parameter for Eq.~(\ref{eq:Psi_sigma}) leads to incorrect results because the uncertainty of $p_0$ does not propagate into the uncertainty of $\varPsi$.
\item Calculating $\Delta \varPsi$ with the lower limit $\underline{p_0}$ that was computed in \S\,\ref{sec:dp} overestimates the uncertainty in $\varPsi$ because the interval $\Delta p_0$ is constructed to be minimal in $p_0$ but not in $\varPsi$ (the upper limit $\overline{p_0}$ underestimates $\Delta \varPsi$, cf. Fig.~\ref{fig:Psi_sigma}). 
\item Transforming $p_0$ to $p$, as in the transformation of the bias (Eq. (\ref{eq:proj})),
\begin{equation}
  \label{eq:proj_Psi}
	\Delta \varPsi(p) = \int_0^{\infty} \Delta \varPsi(p_0) \cdot \rho_{\mathrm{p}_0}(p_0\,|\,p) \,\mathrm{d}p_0
\end{equation} 
averages $\Delta \varPsi$. The result is reasonable, but does not match the definition of credibility intervals in Eq. (\ref{eq:cred_def}).
\end{itemize}  
\noindent
Our idea is to recalculate $\Delta p_0$ credibility intervals with the bivariate probability distribution $\rho(p,\varPsi\,|\,p_0,\varPsi_0)$ of Eq.~(\ref{eq:pdf}). After integrating over $p_0$ and normalizing the bivariate distribution for fixed $p$'s with respect to $\varPsi$, the credibility interval $\Delta\varPsi$ can be computed ($\leftleftarrows$) directly (without loss of generality $\varPsi_0 = 0$):
\begin{eqnarray}
	\rho^*(p, \varPsi)\hspace{-3mm} &=& \hspace{-3mm} \int\limits_{0}^\infty \rho(p,\varPsi\,|\,p_0, 0)\,\mathrm{d} p_0,\\
	\rho(p, \varPsi) \hspace{-3mm} &=& \hspace{-3mm} \frac{\rho^*(p, \varPsi)}{N(p)}  \; ,\,N(p) = \int\limits_{-\pi/2}^{\pi/2} \rho^*(p, \varPsi) \,\mathrm{d}\varPsi, \hspace{4mm}\\
	\Delta\varPsi(p) \hspace{-3mm} &\leftleftarrows& \hspace{-3mm}\int\limits_{-\Delta\varPsi/2}^{\Delta\varPsi/2} \rho(p, \varPsi) \, \mathrm{d} \varPsi = C.
\end{eqnarray}
The numerically computed results are shown in Fig.~\ref{fig:Psi_sigma_cor}. See Tab.~\ref{tab:fig_g} for parametric fitting results.

\section{Simulating confidence intervals}
The following Monte Carlo simulations serve as an illustration of the calculations made in \S\,\ref{sec:bay}.
\label{sec:sim}
\subsection{Degree of polarization}
\label{sec:sim_p}
Our simulation is based on a large number $N$ of $(p_0 / p)$-pairs. For an equally spaced distribution of true signal-to-noise ratios $p_0$, $N_\mathrm{p}$ $p$-values that follow the distribution of Eq. (\ref{eq:dp}) are randomly chosen for each $p_0$ value. Our simulation uses $0 \le p_0 \le 10$ with a step size of $\Delta p_0 = 0.01$ and $N_\mathrm{p} = 200\,000$. That makes $N = 2 \cdot 10^8$ $(p_0 / p)$-pairs in total. To find $(p_0 / p)$-pairs with a specific $p$-value, the continuous distributed pairs in $p$ are binned in the $p$-dimension in intervals of $[p, p+0.01]$, each containing $N_{p_0}$ pairs. 

\begin{figure}[t]
\centering
\includegraphics[width = 1\linewidth]{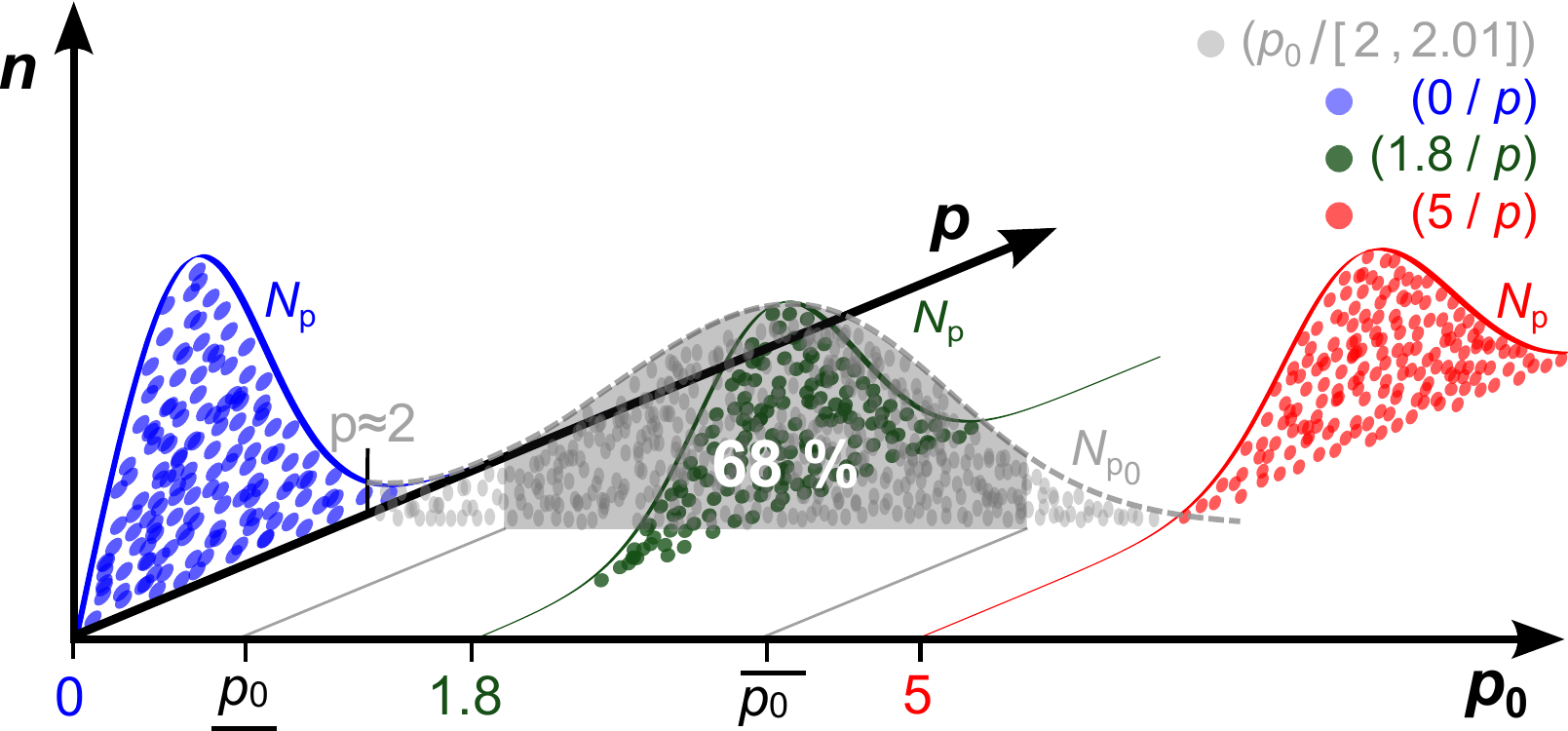}
\caption{Model to explain the calculation of credibility intervals for the degree of polarization with the help of $(p_0 / p)$-pairs: $p$-values are guessed on the basis of uniform distributed $p_0$. The third dimension ($n$) equals the $(p_0 / p)$-pair density. For reasons of clarity only the pairs with $p_0 \in \{0, 1.8, 5\}$ and those with $p = 2$ are shown. For the example $p = 2$, the corresponding $\sigma_1$ credibility interval (gray shaded area) consists of 68\,\% of all pairs $(p_0/2)$ with $p_0 \in [\underline{p_0}, \overline{p_0}]$.}
\label{fig:sim_expl}
\end{figure}

The credibility interval $[\underline{p_0}, \overline{p_0}]$ for each $p$-value can finally be calculated as the narrowest interval in $p_0$ that contains $C \cdot N_\mathrm{p_0}$ data points. Fig. \ref{fig:sim_expl} explains the described method graphically for $p \approx 2$.

\subsection{Angle of polarization}
\label{sec:sim_Psi}
\begin{figure}[t]
\centering
\includegraphics[width = 1\linewidth]{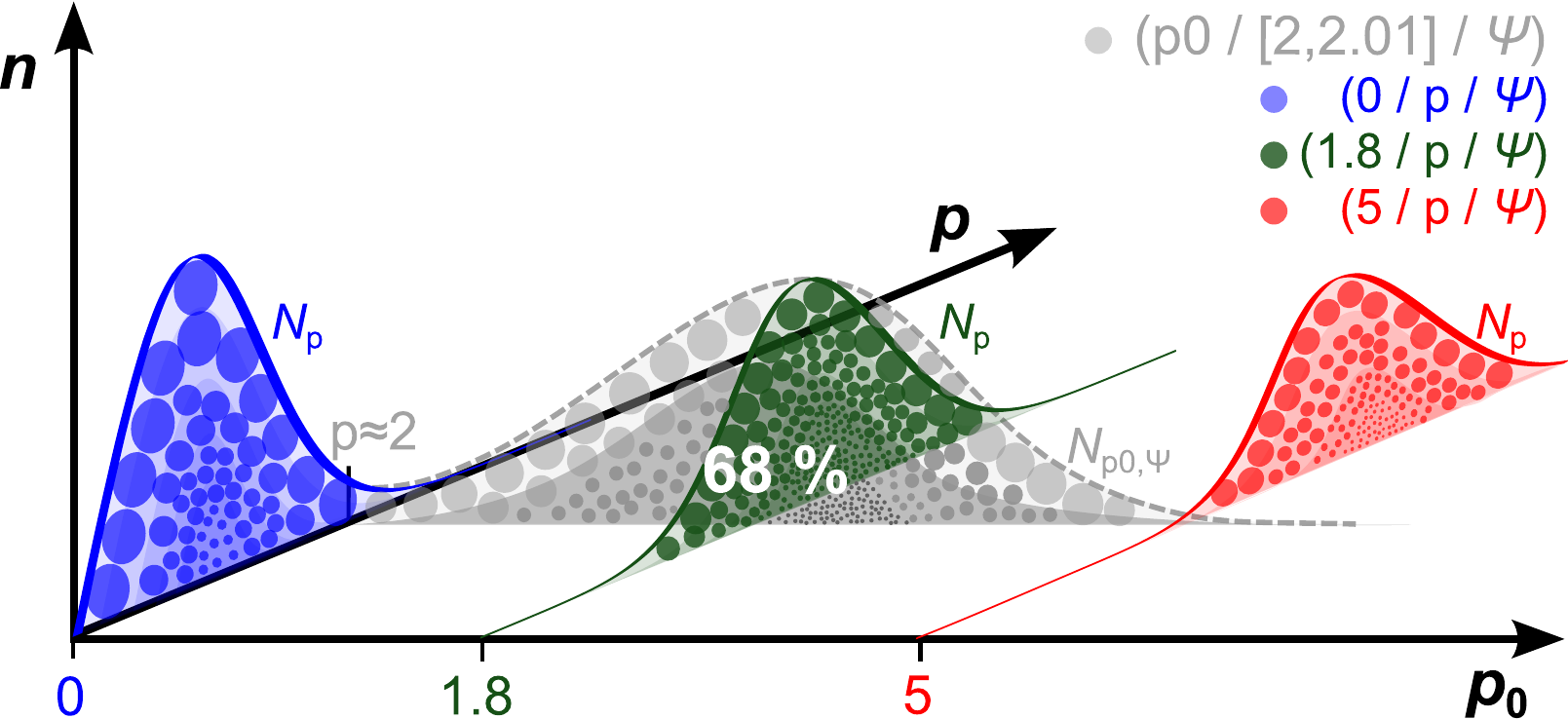}
\caption{Constructing credibility intervals for the polarization angle. The notation is similar to Fig. \ref{fig:sim_expl} but now, each simulated data point is a triplet $(p_0/p/\varPsi)$ plotted against its number density of occurrence $n$. As $\varPsi_0 = 0$, it follows $\Delta \varPsi = \varPsi$. The values of $\varPsi$ are indicated by the size of the dots. For the example $p = 2$, the corresponding $\sigma_1$ credibility interval (gray shaded area) consists of 68\,\% of all triplets $(p_0/2/\varPsi)$ with arbitrary $p_0$ but smallest $\varPsi$. 
}
\label{fig:sim_expl_Psi}
\end{figure}

Simulating interval estimations on the polarization angle ($\Psi_0 = 0$ is assumed) is done in two steps. First, appropriate random data must be simulated:
\begin{enumerate}
\item Choose a true signal-to-noise ratio $p_0 \in [0, 10]$.\vspace{-1mm}
\item Guess a value $p$ following $\rho_\mathrm{p}(p\,|\,p_0)$, cf. Eq. (\ref{eq:dp}).\vspace{-1mm}
\item Guess a value $\varPsi$ following $\rho(p,\!\varPsi\,|\,p_0, 0)$, Eq.~(\ref{eq:pdf}).\vspace{-1mm}
\end{enumerate} 
Repeating these steps numerous times yields a large number of $(p_0 / p / \varPsi)$-triplets. In the second step, analyzing these triplets will result in the desired credibility intervals:
\begin{enumerate}
\item Select all triplets with a specific value of $p \in [p, p+0.01]$, independent of $p_0$ and $\varPsi$.
\vspace{-1mm}
\item Count the selected triplets $\rightarrow N_{\mathrm{p}_0, \Psi}$.
\vspace{-1mm}
\item Sort the triplet list with respect to $\varPsi$.
\vspace{-1mm}
\item Starting counting at $\varPsi = 0$, the credibility interval\footnote{The $\varPsi$-symmetry in Eq. (\ref{eq:pdf}) allows to restrict all calculations on the half credibility interval.}  $\Delta\varPsi/2$ can be obtained as the $\varPsi$ value of the data triplet at list number $C/2 \cdot N_{\mathrm{p_0}, \varPsi}$. 
\end{enumerate}
In this way, the uncertainty in $\varPsi$ can be estimated on the basis of the observed signal to noise ratio $p$.
Repeating these steps for a set of different $p$-values results in the data points plotted in Fig. \ref{fig:Psi_sigma_cor}. Again, the method used is explained graphically in Fig. \ref{fig:sim_expl_Psi}.

\section{Results}
\label{sec:res}
\subsection{Degree of polarization}
\label{sec:degofpol}
Based on the results shown in Fig.~\ref{fig:offset_p}, the best estimator $\hat{p}_0$ can be chosen by means of lowest expected bias. The excellent results of the approximation $\hat{p}_\mathrm{0,\,ebal} = (p^2 -1)^{0.5}$  for $p > 2.8$ in combination with its analytical form makes this estimator a good choice for high signal-to-noise ratios $p$. The regions of the best performance and least square fits for $\hat{p}_0$ within those regions are the following:
\begin{eqnarray}
\label{eq:para_p0_mlstart}
\hat{p}_\mathrm{0,\,ml} \hspace{2.0mm} = 0  \hspace{36.5mm}    p \in [0, \sqrt{2}] \hspace{0.3mm}\\
\label{eq:para_p0_ml}
\nonumber \hat{p}_\mathrm{0,fit\,ml} \approx  (p\!-\!\!\!\sqrt{2})^{0.4542}\!\!+(p\!-\!\!\!\sqrt{2})^{0.4537}\!+ (p\!-\!\!\!\sqrt{2})/4\\
\Delta = [-0.0078, 0.011], \;\;\hspace{1.7mm} p \in [\!\sqrt{2}, 1.70] \\ 
\nonumber \hat{p}_\mathrm{0,fit\,mean} \hspace{-0.4mm}\approx\hspace{-0.3mm} 22 p^{0.11} - 22.076\hspace{29.3mm}\\
\Delta = [-0.0039, 0.0035],  p \in [1.70, 2.23]\\
\label{eq:para_p0_med}
\nonumber \hat{p}_\mathrm{0,fit\,med} \hspace{0.7mm}\approx  1.8 p^{0.76} - 1.328 \hspace{29.9mm}\\
\Delta = [-0.0024, 0.002], \hspace{1.7mm} p \in [2.23, 2.83]\\ 
\label{eq:para_p0_ebal}
\hat{p}_\mathrm{0,\,ebal} \hspace{3mm} = (p^2 -1)^{0.5} \hspace{19.0mm}p \in [2.83, \infty]
\end{eqnarray}

\noindent
The Eq. (\ref{eq:para_p0_mlstart})-(\ref{eq:para_p0_ebal}) are not one combined estimator for all values of $p$, but a collection of different estimators, each working in the region of best expected performance. Therefore, discontinuous jumps at the interval edges are not a lack of accuracy, but a result of different estimators. 
The motivation for the functional forms in Eq. (\ref{eq:para_p0_ml}) - (\ref{eq:para_p0_med}) and in Eq. (\ref{eq:fit_f}) and (\ref{eq:fit_g}) are not physically driven but determined by mathematical intuition and the best fitting results. $\Delta = \hat{p}_{0,\mathrm{fit}\,e} - \hat{p}_{0,\,e}$ indicates the range of maximal deviation between the fitted curves and the respective estimator $e$.

Concerning the construction of interval estimations, confidence and credibility intervals show very similar results for $p > 6$. At low signal-to-noise ratios $p < 3$ they differ significantly. The choice of which method to use depends on the question which shall be answered: is it the chance $C$ that the confidence interval includes the true parameter, or the probability $C$ that the true parameter is within the credibility interval?

Lower and upper limits for $\sigma_1$, $\sigma_2$, and $\sigma_3$ confidence intervals are presented in Fig. \ref{fig:CI}. A functional description for the tangential construction method is obtained by fitting the function
\begin{equation}
  \label{eq:fit_f}
	f(p) = A p ^{B} - C p^{-D} + E p
\end{equation}
to the computed data points with the least square method. The results are listed in Table \ref{tab:fit_f}.

\begin{table}[h]
\centering
\caption{Fitting results of Eq.\,(\ref{eq:fit_f}) for the lower and upper confidence interval limits for the degree of polarization. See also Fig. \ref{fig:CI}. The deviation between the numerically calculated data points and $f$ is maximal $\pm 0.05$.  $f = 0$ for all undefined regions with $p < 6$.}
\vspace{0.5mm}
\begin{small}
\begin{tabularx}{0.46\textwidth}{p{4.7mm}p{6.6mm}p{6.6mm}p{2mm}p{4.5mm}p{6.0mm}p{19mm}}
\toprule
   $f$               & $A$ & $\hspace{-1.5mm}B$ & $\hspace{-3mm}C$ & $D$ & $E$ & $\hspace{4mm}$validity \vspace{1mm}\\
   \toprule
$\underline{p_1}(p)$ & -1.17 & $\hspace{-1.5mm}0.027$ & $\hspace{-3mm}5413 $ & 23.63 & 1.018 & $\hspace{0mm}1.51 \le p\le 6$ \vspace{1mm}\\
$\underline{p_2}(p)$ & -1.86 & $\hspace{-1.5mm}0.192$ & $\hspace{-3mm}1.0e5$ & 13.58 & 1.083 & $\hspace{0mm}2.50 \le p\le 6$ \vspace{1mm}\\
$\underline{p_3}(p)$ & -2.72 & $\hspace{-1.5mm}0.085$ & $\hspace{-3mm}1.0e7$ & 13.94 & 1.018 & $\hspace{0mm}3.41 \le p\le 6$ \vspace{1mm}\\
$\overline{p_1}(p)$  & 1.691 & $\hspace{-1.5mm}1.000$ & $\hspace{-3mm}0.000$ & 0.000 & 0.000 & $\hspace{0mm}0 \le p \le 0.77$ \vspace{1mm}\\
                     & 0.715 & $\hspace{-1.5mm}0.328$ & $\hspace{-3mm}0.021$ & 4.766 & 0.936 & $\hspace{0mm}0.77 < p \le 6$\vspace{1mm}\\
$\overline{p_2}(p)$  & 5.441 & $\hspace{-1.5mm}1.000$ & $\hspace{-3mm}0.000$ & 0.000 & 0.000 & $\hspace{0mm}0 \le p \le 0.22$ \vspace{1mm}\\
                     & 1.741 & $\hspace{-1.5mm}0.083$ & $\hspace{-3mm}0.027$ & 1.913 & 0.980 & $\hspace{0mm}0.22 < p \le 6$\vspace{1mm}\\
$\overline{p_3}(p)$  & 22.86 & $\hspace{-1.5mm}1.000$ & $\hspace{-3mm}0.000$ & 0.000 & 0.000 & $\hspace{0.0mm}0\!\le\!p\!\le 0.065$ \vspace{1mm}\\
                     & 4.621 & $\hspace{-1.5mm}-0.19$ & $\hspace{-3mm}1.937$ & 0.435 & 1.091 & $\hspace{0.0mm}0.065\!<\! p\! \le 6$ \vspace{1mm}\\
\bottomrule
\end{tabularx}
\label{tab:fit_f}
\end{small}
\end{table}

\noindent
Calculated lower and upper limits for of $\sigma_1$, $\sigma_2$, and $\sigma_3$ credibility intervals are presented in Fig. \ref{fig:CI_bay}. Functional descriptions are obtained as before. The results are listed in Table \ref{tab:fit_f_bay}. For high signal-to-noise ratios $p > 6$, the Gaussian approximation can be used.
\begin{table}[h]
\centering
\caption{Fitting results of Eq.\,(\ref{eq:fit_f}) for the lower and upper credibility interval limits for the degree of polarization. See also Fig. \ref{fig:CI_bay}. The deviation between the numerically calculated data points and the fit $f$ is maximal $\pm 0.025$. $p < 6$ for all cases and $f = 0$ for all remaining undefined regions.}
\vspace{0.5mm}
\begin{small}
\begin{tabularx}{0.46\textwidth}{p{5.1mm}p{5.6mm}p{8.2mm}p{6.8mm}p{3.7mm}p{8.2mm}p{19mm}}
\toprule
   $f$               & $A$ & $B$ & $\hspace{-1.8mm}C$ & $\hspace{-2.5mm}D$ & $E$ & \hspace{-0.9mm}validity \vspace{1mm}\\
   \toprule
         $\underline{p_1}(p)$ & 4.241 & 1.021 & \hspace{-1.8mm}2.286 & $\hspace{-2.5mm}1.134 $ & -3.535 & \hspace{-0.9mm}$p \ge 1.72$ \vspace{1mm}\\
         $\underline{p_2}(p)$ & 0.468 & 1.177 & \hspace{-1.8mm}3.974 & $\hspace{-2.5mm}0.874 $ & 0.145  & \hspace{-0.9mm}$p \ge 2.54$ \vspace{1mm}\\
         $\underline{p_3}(p)$ & 1.327 & 1.121 & \hspace{-1.8mm}7.599 & $\hspace{-2.5mm}1.131 $ & -1.00  & \hspace{-0.9mm}$p \ge 3.45$ \vspace{1mm}\\
         $\overline{p_1}(p)$  & 0.292 & 2.063 & \hspace{-1.8mm}-1.00 & $\hspace{-2.5mm}0.000 $ & 0.000  & \hspace{-0.9mm}$p \le 1.72$ \vspace{1mm}\\
         											& 0.855 & 0.020 & \hspace{-1.8mm}17.87 & $\hspace{-2.5mm}6.012 $ & 1.009  & \hspace{-0.9mm}$p > 1.72$ \vspace{1mm}\\
         $\overline{p_2}(p)$  & 1.819 & 1.185 & \hspace{-1.8mm}-2.00 & $\hspace{-2.5mm}0.000 $ & -1.345 & \hspace{-0.9mm}$p \le 2.54$ \vspace{1mm}\\
         											& 1.910 & -0.028& \hspace{-1.8mm}13794 & $\hspace{-2.5mm}11.21 $ & 1.018  & \hspace{-0.9mm}$p > 2.54$ \vspace{1mm}\\
         $\overline{p_3}(p)$  & 0.564 & 1.632 & \hspace{-1.8mm}-3.00 & $\hspace{-2.5mm}0.000 $ & 0.000  & \hspace{-0.9mm}$p \le 1.40$ \vspace{1mm}\\ 
      										   	& 1.058 & 1.000 & \hspace{-1.8mm}-2.47 & $\hspace{-2.5mm}0.000 $ & 0.00   & \hspace{-4.9mm}$1.4 \!<\! p \!\le\! 3.5$ \vspace{1mm}\\ 
         											& 4.130 & -0.38 & \hspace{-1.8mm}2.0e5 & $\hspace{-2.5mm}10.65 $ & 1.140  & \hspace{-0.9mm}$p > 3.50$ \vspace{1mm}\\
\bottomrule
\end{tabularx}
\label{tab:fit_f_bay}
\end{small}
\end{table}
   
\subsection{Angle of polarization}
\label{sec:angofpol}
Despite its symmetrical probability distribution function $\rho_\Psi$, constructing confidence intervals for the angle of polarization is  not trivial because of its dependency on $p_0$. Calculated lower and upper limits for $\sigma_1$, $\sigma_2$, and $\sigma_3$ credibility intervals are presented in Fig.~\ref{fig:Psi_sigma_cor}. A functional description is obtained by fitting the function
\begin{equation}
	\label{eq:fit_g}
	g(p) = A \, \Big(B + \tanh{\big(C\left(D-p\right)\big)}\Big)-E p  \hspace{2mm} \big[\textnormal{in deg.}\big]
\end{equation}
to the computed data points with the least square method.
\begin{table}[t]
\centering
\vspace{-2mm}
\caption{Fitting results of Eq.\,(\ref{eq:fit_g}) for the lower and upper credibility interval limits for the angle of polarization. See also Fig. \ref{fig:Psi_sigma_cor}. The deviation between the numerically calculated data points and $g$ is maximal $\pm 1.7^\circ$.}
\vspace{1mm}
\begin{small}
\begin{tabularx}{0.46\textwidth}{p{4.7mm}p{5.5mm}p{5.5mm}p{5.5mm}p{5.5mm}p{5.5mm}p{19mm}}
	\toprule
	$g$    & $A$ & $B$ & $C$ & $D$ & $E$ & validity \vspace{1mm}\\
	\toprule
	$\sigma_1(p)$ & 32.50 & 1.350 & 0.739 & 0.801 & 1.154 & $0.0\!< \!p\! \le\! 6.0$  \vspace{1mm}\\
	$\sigma_2(p)$ & 65.65 & 0.323 & 0.858 & 2.688 & 0.000 & $0.0\! < \!p\! \le \!2.2$ \vspace{1mm}\\
                & 517   & 1.044 & 0.806 & 0.015 & 2.186 & $2.2 \!< \!p \!\le \!6.0$ \vspace{1mm}\\
	$\sigma_3(p)$ & 62.88 & 0.423 & 1.385 & 3.546 & 0.000 & $0.0 \!<\! p\! \le\! 3.2$  \vspace{1mm}\\
              & 102     & 1.380 & 1.327 & 2.506 & 3.958 & $3.2 \!<\! p \!\le\! 6.0$ \vspace{1mm}\\
  \bottomrule
\end{tabularx}
\label{tab:fig_g}
\end{small}
\end{table}

\noindent
For $p>6$, a very simple description can be obtained using Gaussian error propagation on Eq.~(\ref{eq:dop_stokes}) and (\ref{eq:psi_stokes}) with $\sigma_\mathrm{u} = \sigma_\mathrm{q} = \sigma$ \citep[][Eq. (70)]{Serkowski1962289}:

\begin{eqnarray}
	\hspace{0mm}\sigma_1(p) &=& 28.65^\circ/p \hspace{1.5cm} p > 6\hspace{2.5mm} \\
	\hspace{0mm}\sigma_2(p) &=& 57.30^\circ/p \hspace{1.5cm} p > 6\hspace{2.5mm} \\
	\hspace{0mm}\sigma_3(p) &=&  85.95^\circ/p \hspace{1.5cm} p > 6.\hspace{2.5mm}
\end{eqnarray}
\newpage
\section{Conclusions}
\label{sec:con}
Polarimetric measurements incorporate a non-Gaussian statistic in terms of the degree and the angle of polarization. The aim of this work was to present a systematic overview of the statistics that are necessary to analyze such measurements. In particular, we calculated point estimations for the degree of polarization and interval estimations for the angle of polarization with a Bayesian approach for the first time.

From an observational point of view, the Bayesian analysis shows substantial advantages compared to frequentist analysis. It allows direct access for the best estimator to be chosen on the basis of observational data and produces interval estimations with a meaningful interpretation.

In conclusion, observational polarimetric data can be recalculated in terms of signal-to-noise ratios $p$ using Eq. (\ref{eq:p}). The choice of the best estimator, the best estimated value~$\hat{p}_0$, confidence and credibility intervals for the degree of polarization, and credibility intervals for the angle of polarization can then be obtained directly with the approximated formulas in \S\,\ref{sec:degofpol} and \S\,\ref{sec:angofpol}. Using Eq. (\ref{eq:p}) reversed, all calculated values for the signal-to-noise ratio of the degree of polarization can be expressed as degree of polarization.

\enlargethispage{-26.5ex} 
\begin{acknowledgements}
This work is supported by the {\it Bundesministerium f\"{u}r Wirtschaft und Technologie} through  the {\it Deutsches Zentrum f\"{u}r Luft- und Raumfahrt e.V.} under the grant number 50OO1110 and 50QR1101.
We want to thank Prof. M. Zerner for discussing mathematical issues concerning confidence intervals.
\end{acknowledgements}

\begin{small}
\bibliographystyle{aa} 
\bibliography{lit}
\end{small}
\enlargethispage{-26.5ex} 
\end{document}